\documentclass[aps,twocolumn,prd,groupedaddress,nofootinbib,amssymb,eqsecnum,epsfig]{revtex4-1}
\usepackage{graphicx}
\usepackage{bm}
\usepackage{amsmath}
\usepackage{color}
\usepackage{amsfonts}

\begin{document}
\newcommand{\newc}{\newcommand}

\newcommand{\ben}{\begin{eqnarray}}
\newcommand{\een}{\end{eqnarray}}
 \newcommand{\lb}{\label}
\newc{\be}{\begin{equation}}
\newc{\ee}{\end{equation}}
\newc{\ba}{\begin{eqnarray}}
\newc{\ea}{\end{eqnarray}}
\newc{\bea}{\begin{eqnarray*}}
\newc{\eea}{\end{eqnarray*}}
\newc{\D}{\partial}
\newc{\ie}{{\it i.e.} }
\newc{\eg}{{\it e.g.} }
\newc{\etc}{{\it etc.} }
\newc{\etal}{{\it et al.}}
\newcommand{\nn}{\nonumber}
\newc{\ra}{\rightarrow}
\newc{\lra}{\leftrightarrow}
\newc{\lsim}{\buildrel{<}\over{\sim}}
\newc{\gsim}{\buildrel{>}\over{\sim}}

\title{Odd-parity stability of black holes in Einstein-Aether gravity}

\author{
Shinji Tsujikawa$^{1}$, Chao Zhang$^{2,3,4}$, Xiang Zhao$^{2}$, and Anzhong Wang$^{2}$}

\affiliation{$^1$Department of Physics, Waseda University, 3-4-1 Okubo, 
Shinjuku, Tokyo 169-8555, Japan\\
$^2$GCAP-CASPER, Physics Department, Baylor University, Waco, Texas, 76798-7316, USA\\
$^3$Institute for Theoretical Physics \& Cosmology, Zhejiang University of Technology, Hangzhou, 310023, China\\
$^4$United Center for Gravitational Wave Physics (UCGWP),
Zhejiang University of Technology, Hangzhou, 310023, China
}

\date{\today}

\begin{abstract}

In Einstein-Aether theory, we study the stability of black holes against 
odd-parity perturbations on a spherically symmetric and static background. 
For odd-parity modes, there are two dynamical degrees of freedom 
arising from the tensor gravitational sector and Aether vector field. 
We derive general conditions under which neither ghosts nor Laplacian instabilities 
are present for these dynamical fields. We apply these results to concrete black 
hole solutions known in the literature and show that some of those solutions 
can be excluded by the violation of stability conditions. 
The exact Schwarzschild solution present for $c_{13} = c_{14} = 0$, 
where $c_i$'s are the four coupling constants 
of the theory with $c_{ij}=c_i + c_j$, is prone to Laplacian instabilities along 
the angular direction throughout the horizon exterior.
However, we find that the odd-parity instability of high radial and angular 
momentum modes is absent for black hole solutions with 
$c_{13}  = c_4 = 0$ and $c_1 \geq 0$.

\end{abstract}

\pacs{04.50.Kd,95.30.Sf,98.80.-k}

\maketitle

\section{Introduction}

General Relativity (GR) is a fundamental theory of gravity well tested 
by solar-system experiments. With the dawn of gravitational-wave 
astronomy, it is now possible to probe the validity of GR 
around black holes (BHs) and neutron stars \cite{Abbott2016,GW170817}.
Recently, there has been growing interest 
in searching for extra degrees of freedom beyond GR and 
standard model of particle physics in such 
a strong gravity regime \cite{Berti,Barack}. 
The existence of new degrees of freedom is also motivated by the 
firm observational evidence of dark matter and 
dark energy \cite{Bertone:2004pz,Copeland:2006wr,Clifton:2011jh}.

The construction of GR is based on Lorentz invariance (LI), which 
is a continuous symmetry invariant under the 4-dimensional diffeomorphism. 
In discrete spacetime that can arise from the quantization of gravity, 
the Lorentz symmetry can be broken at very high energy. 
The violation of LI in standard model fields 
is tightly limited from various experiments \cite{Kostelecky:2008ts,Mattingly:2005re}, 
but the Lorentz violation in the gravity sector 
is much less constrained \cite{Flowers:2016ctv,Bourgoin:2017fpo}. 
Ho\v{r}ava gravity \cite{Horava:2009uw,Blas:2009qj} is an example 
of allowing for gravitational Lorentz violation 
at high energy, in which a Lifshitz-type anisotropic scaling is introduced 
to realize a power-counting renormalizable theory of 
gravity (for a recent review of 
Ho\v{r}ava gravity, see, for example, \cite{Wang:2017brl} and references therein).

There is yet the other type of a gravitational Lorentz-violating scenario 
dubbed Einstein-Aether theory \cite{Jacobson:2000xp,Jacobson}.
In this scenario there is a unit time-like vector (Aether) field $u^{\alpha}$ 
at every point in spacetime characterized 
by the metric tensor $g_{\alpha \beta}$, so it breaks local Lorentz symmetry under a rotation.
This is a subclass of vector-tensor theories possessing two derivative terms 
of the Aether field. The existence of a unit Aether field is ensured by the
constraint  $g_{\alpha \beta} u^{\alpha} u^{\beta}=-1$ [with the metric 
signature $(-,+,+,+)$], 
which appears as the Lagrange multiplier 
$\lambda (g_{\alpha \beta} u^{\alpha} u^{\beta}+1)$ in the action. 
We note that generalized Proca theories with a broken $U(1)$ gauge 
symmetry \cite{Heisenberg,Tasinato,Allys,Jimenez2016,Heisenberg:2016eld} 
do not have such a constraint, so the vector-field dynamics 
is generally different from that in Einstein-Aether theory.

In Einstein-Aether theory there are scalar, transverse vector, 
and tensor perturbations, whose propagation 
speeds $c_S$, $c_V$, $c_T$ on the Minkowski background 
are generally different from that of light \cite{Jacobson:2004ts}. 
To ensure the stability of Minkowski spacetime, we require that all
of $c_S^2$, $c_V^2$, and $c_T^2$ are positive.
Moreover, the observations of gravitational Cerenkov 
radiation \cite{Elliott:2005va}, solar system tests \cite{Foster:2005dk}, 
big-bang nucleosynthesis \cite{Carroll:2004ai}, 
binary pulsars \cite{Foster:2007gr,Yagi:2013ava}, and gravitational 
waves \cite{Gong:2018cgj,Oost:2018tcv} 
put constraints on the dimensionless 
coupling constants $c_{1,2,3,4}$ of Aether derivative interactions. 
In particular, the gravitational-wave event GW170817 \cite{GW170817} together with the 
gamma-ray burst 170817A \cite{Goldstein} placed the upper limit $|c_T-1| \lesssim 10^{-15}$, 
which translates to $|c_{13}| \lesssim 10^{-15}$ \cite{Gong:2018cgj,Oost:2018tcv}, 
where $c_{ij}:=c_{i} + c_j$.
However, there are still theoretically viable parameter spaces 
in which all the observational constraints are satisfied.

In Einstein-Aether theory, the existence and properties of 
spherically symmetric vacuum solutions have been 
extensively studied in the literature \cite{Eling:2006ec,Eling:2007xh,Tamaki:2007kz,Barausse:2011pu,Blas:2011ni,Berglund:2012bu,
Gao:2013im,Lin:2014eaa,Ding:2015kba,Ding:2016wcf,Ding:2018whp,Chan:2019mdn,Chan:2020amr,
Khodadi:2020gns}. 
Some of them were already excluded by the combination of 
observational bounds mentioned above.  
However, the recent papers \cite{Zhang:2020too,Oost:2021tqi} have shown 
the presence of spherically symmetric and static BH solutions compatible 
with current observational constraints.
Since the speeds of scalar and transverse vector perturbations 
can be arbitrarily large, there exists a universal horizon corresponding 
to a causal boundary of any large speeds of 
propagation \cite{Blas:2011ni,Berglund:2012bu,Wang:2017brl}. 
The universal horizon can exist inside the event horizon, so that 
particles can cross the event horizon to escape toward infinity.
It is expected that this unique feature of Einstein-Aether BHs may 
leave some distinguished signatures in the gravitational-wave measurements 
of binary BHs.

In this paper, we study the stability of spherically symmetric and static BHs 
against odd-parity perturbations in Einstein-Aether theory.  
We first identify two dynamical gauge-invariant perturbations corresponding 
to the tensor and vector propagations. 
Then, we obtain the second-order action of odd-parity perturbations 
and explicitly derive stability conditions for the absence 
of ghosts and Laplacian instabilities. 
The tensor and vector propagation speeds along the radial and angular 
directions are different from those in Minkowski spacetime.
Thus, our analysis of BH perturbations in the odd-parity sector 
provides new stability conditions for Einstein-Aether BHs. 
We also note that our general formulation of odd-parity perturbations 
will be useful to study the propagation of gravitational waves during the inspiral 
and ringdown phases of binary BHs.

We apply our conditions to the Einstein-Aether BH solutions 
known in the literature. We show that an exact Schwarzschild BH present 
for the couplings $c_{13} =0$ and $c_{14} =0$ is excluded by
the Laplacian instability along the angular direction. 
The BH solutions with $c_{13} =0$, $c_{14} \neq 0$, and $c_4 \neq 0$ 
are prone to the ghost instability by imposing a superluminal propagation 
of the transverse vector mode ($c_4<0$) to avoid the gravitational Cerenkov radiation.
However,  provided that $c_1 \geq 0$, the BH solutions with $c_{13} =0$ and $c_4=0$ 
are stable against odd-parity perturbations with high radial and angular momentum modes.
Thus, our general stability conditions are sufficiently powerful 
to distinguish between unstable and stable BHs in Einstein-Aether theory.

\section{Background equations of motion}
\label{backsec}

We begin with the Einstein-Aether theory given 
by the action \cite{Jacobson:2004ts}
\be
{\cal S}=\frac{1}{16\pi G_{\ae}} \int  \sqrt{-g}\,{\rm d}^4 x 
\left[ R+{\cal L}_{\ae}+\lambda (g_{\alpha \beta} u^{\alpha} u^{\beta}+1) 
\right],
\label{action}
\ee
where $G_{\ae}$ is a constant, $R$ is the Ricci scalar, $g$ is the determinant of 
metric tensor $g_{\alpha \beta}$, $\lambda$ is a Lagrange multipler, 
$u^{\alpha}$ is the Aether vector field, and 
\be
{\cal L}_{\ae}=-{M^{\alpha \beta}}_{\mu \nu}
\nabla_{\alpha} u^{\mu} \nabla_{\beta} u^{\nu}\,,
\ee
with 
\be
{M^{\alpha \beta}}_{\mu \nu}:=c_1 g^{\alpha \beta} g_{\mu \nu}
+c_2 \delta^{\alpha}_{\mu} \delta^{\beta}_{\nu}
+c_3 \delta^{\alpha}_{\nu} \delta^{\beta}_{\mu}
-c_4 u^{\alpha} u^{\beta} g_{\mu \nu}\,.
\ee
Here, the Greek indices represent from 0 to 3, 
$\nabla_{\alpha}$ is a covariant derivative operator 
with respect to the metric tensor $g_{\mu \nu}$, and $c_i$'s 
are four dimensionless coupling constants.

Variation of the action (\ref{action}) with respect to $\lambda$
leads to
\be
u^{\alpha} u_{\alpha}+1=0\,.
\label{be1}
\ee
This constraint ensures the existence of a time-like unit vector field, 
so that there is a preferred frame responsible for the breaking of LI.
Varying Eq.~(\ref{action}) with respect to $u^{\mu}$, 
it follows that 
\be
\nabla_{\mu} {J^{\mu}}_{\alpha}+\lambda u_{\alpha}
+c_4 \dot{u}^{\mu} \nabla_{\alpha} u_{\mu}=0\,,
\label{be2}
\ee
where
\ba
{J^{\mu}}_{\alpha} &:=& 
{M^{\mu \nu}}_{\alpha \beta} 
\nabla_{\nu} u^{\beta}\,,\label{Jmu} \\
\dot{u}^{\mu} &:=& 
u^{\beta} \nabla_{\beta} u^{\mu}\,.
\ea
Multiplying Eq.~(\ref{be2}) by $u^{\alpha}$ and using Eq.~(\ref{be1}), 
the Lagrange multiplier can be expressed as 
\be
\lambda=u^{\alpha} \nabla_{\mu} {J^{\mu}}_{\alpha}
+c_4 \dot{u}^{\mu} \dot{u}_{\mu}\,. 
\label{Lagm}
\ee

For the general line element ${\rm d}s^2=g_{\mu \nu}{\rm d}x^{\mu}{\rm d} x^{\nu}$, 
the gravitational field equations derived by the variation of 
(\ref{action}) with respect to $g_{\mu \nu}$ are 
\ba
\hspace{-0.5cm}
G_{\alpha \beta}
&=&\nabla_{\mu} \left[ u_{(\alpha} {J^{\mu}}_{\beta )}
+u^{\mu}J_{(\alpha \beta)}-u_{(\alpha} {J_{\beta)}}^{\mu} \right]
\nonumber \\
\hspace{-0.5cm}
&&+ c_1 \left( \nabla_{\alpha} u^{\nu} \nabla_{\beta} u_{\nu} 
-\nabla^{\nu} u_{\alpha} \nabla_{\nu} u_{\beta} \right)
\nonumber \\
\hspace{-0.5cm}
&&
+c_4 \dot{u}_{\alpha} \dot{u}_{\beta}+\frac{1}{2} g_{\alpha \beta}{\cal L}_{\ae}
+\lambda u_{\alpha} u_{\beta}\,,
\label{Ein}
\ea
where $G_{\alpha \beta}$ is the Einstein tensor.

In general, the theory admits three different species of gravitons, the spin-0, spin-1, 
and spin-2 ones. 
According to the perturbative analysis on the Minkowski background, 
their squared speeds are given by \cite{Jacobson:2004ts}
\ba
c_S^2 & = & \frac{c_{123}(2-c_{14})}{c_{14}(1-c_{13}) (2+c_{13} + 3c_2)}\,,\label{cSm}\\
c_V^2 & = & \frac{2c_1 -c_{13} (2c_1-c_{13})}{2c_{14}(1-c_{13})}\,,\label{cVm}\\
c_T^2 & = & \frac{1}{1-c_{13}}\,,\label{cTm}
\ea
where  $c_{ijk}:=c_i + c_j + c_k$, and $c_{S,V, T}$ represent the speeds of the spin-0, spin-1, 
and spin-2 gravitons, respectively. 
If we require that the theory:  (i)  be self-consistent, such as free of ghosts and instability; 
and (ii) be compatible with all the observational constraints obtained so far, 
it was found that the parameters $c_{i}$'s 
must satisfy the conditions \cite{Oost:2018tcv}
\ba
\label{CD1a}
&& \left|c_{13}\right| \lesssim 10^{-15},\\
\label{CD1b}
&& 0 < c_{14} \leq 2.5 \times 10^{-5},\\
\label{CD1c}
&&  c_{14} \leq c_2 \leq  0.095,\\
\label{CD1d}
&&  c_{4} \leq 0.
\ea
It should be noted that the above conditions assure $c_{S,V, T} \ge 1$, that is, 
all the propagation speeds are not subluminal, in order to avoid the 
gravitational Cerenkov radiation \cite{Elliott:2005va}. 
Later, we shall come to this point again when we study the odd-parity 
stability of BHs in Sec.~\ref{BHstasec}.

With the above in mind, let us consider a spherically symmetric and static 
background given by 
\be
{\rm d}s^{2} =-f(r) {\rm d}t^{2} +h^{-1}(r) {\rm d}r^{2} + 
r^{2} \left( {\rm d}\theta^{2}+\sin^{2}\theta\, 
{\rm d} \varphi^{2} \right)\,,
\label{metric}
\ee
where $f$ and $h$ depend on the distance $r$ from 
the center of symmetry.
The Aether-field profile compatible with the 
background (\ref{metric}) is of the form 
\be
u^{\mu}=\left( a(r), b(r), 0, 0 \right)\,,
\label{vector_ansatz}
\ee
where $a$ and $b$ are functions of $r$. 
The constraint (\ref{be1}) gives the following relation 
\be
b=\epsilon \sqrt{(a^2 f -1)h}\,,
\label{bso}
\ee
where $\epsilon= \pm 1$. 
The existence of the Aether-field profile (\ref{bso}) requires that 
$(a^2 f -1)h \geq 0$.

Under the constraint (\ref{bso}), there are three independent 
background equations of motion following from (\ref{be2}) 
and (\ref{Ein}), with the Lagrange multiplier $\lambda$ 
determined by  Eq.~(\ref{Lagm}). 
Then, the $\alpha=0$ component of Eq.~(\ref{be2}) 
and $(\alpha, \beta)=(1,1), (2,2)$ components of 
Eq.~(\ref{Ein}) lead to 
\ba
& &
2h \alpha_1 a''+\alpha_1 h' a'+2h \alpha_2 f''+\alpha_2 f' h'
+\alpha_3 a'^2+\alpha_4 f'^2 \nonumber \\
& &+\alpha_5 a' f' +\alpha_6 f'+\alpha_7 h'+\alpha_8 a'+\alpha_9=0\,,
\label{back1}\\
& &
2h \beta_1 a''+\beta_1 h' a'+2h \beta_2 f''+\beta_2 f' h'
+\beta_3 a'^2+\beta_4 f'^2 \nonumber \\
& &+\beta_5 a' f'+\beta_6 f'+\beta_7 h'+\beta_8 a'+\beta_9=0\,,
\label{back2}\\
& &
2h \mu_1 a''+\mu_1 h' a'+2h \mu_2 f''+\mu_2 f' h'
+\mu_3 a'^2+\mu_4 f'^2 \nonumber \\
& &+\mu_5 a' f'+\mu_6 f'+\mu_7 h'+\mu_8 a'=0\,,
\label{back3}
\ea
where a prime represents the derivative with respect to $r$. 
The explicit form of $\lambda$ as well as the coefficients $\alpha_1, \cdots \alpha_9$, 
$\beta_1, \cdots, \beta_9$, and $\mu_1, \cdots \mu_8$
are given in Appendix A. 
We note that Eqs.~(\ref{back1})-(\ref{back3}) hold irrespective 
of the sign of $\epsilon$ in Eq.~(\ref{bso}).
For given coupling constants $c_i$'s, 
the variables $f$, $h$, and $a$ are known by integrating
Eqs.~(\ref{back1})-(\ref{back3}) with appropriate 
boundary conditions. 

\section{Second-order action of odd-parity perturbations 
and general stability conditions}
\label{oddparitysec}

In this section, we derive the second-order action of dynamical 
perturbations in the odd-parity sector to study the stability of 
spherically symmetric and static BH solutions in Einstein-Aether theory. 
Analogous to the analysis performed in Ref.~\cite{Kase:2018voo} 
in the context of generalized Proca theories, we consider metric perturbations 
$h_{\mu \nu}$ on the background 
(\ref{metric}) as well as the perturbation of the Aether field. 
We express the perturbations in terms of the sum of spherical 
harmonics $Y_{lm} (\theta, \varphi)$.

For $l \geq 2$, we choose the Regge-Wheeler gauge in which 
the components $h_{ij}$, where $i$ and $j$ correspond to either 
$\theta$ or $\varphi$, 
vanish \cite{Regge:1957td,Zerilli:1970se}. 
For the dipole ($l=1$),  the metric components $h_{ij}$ vanish 
identically, so we need to handle this case separately.
In the following, we first study the case $l \ge 2$ and then 
proceed to the discussion for $l=1$.

\subsection{$l \geq 2$}

In the Regge-Wheeler gauge, the nonvanishing 
components of metric perturbations are given by 
\ba
h_{ti} &=& \sum_{l,m} Q_{lm} (t,r) E_{ij} \partial^j Y_{lm} 
(\theta, \varphi)\,,\\
h_{ri} &=& \sum_{l,m} W_{lm} (t,r) E_{ij} \partial^j Y_{lm} 
(\theta, \varphi)\,,
\ea
where the subscripts $i$, $j$ represent either $\theta$ 
or $\varphi$ with the notation 
$\partial^j Y_{lm}=\partial Y_{lm}/\partial x_j$, and 
$Q_{lm}$ and $W_{lm}$ are functions of $t$ and $r$.
The tensor $E_{ij}$ is defined by $E_{ij}=\sqrt{\gamma}\, \varepsilon_{ij}$, 
where $\gamma=\sin^2 \theta$ is the determinant of two dimensional 
metric $\gamma_{ij}$ on the sphere and $\varepsilon_{ij}$ is the 
anti-symmetric symbol with $\varepsilon_{\theta \varphi}=1$. 

In the presence of odd-parity perturbations, the covariant Aether field 
is expressed as 
\be
u_{\mu}=\left( -a(r) f(r), \frac{b(r)}{h(r)}, u_{\theta}, u_{\varphi} \right)\,,
\label{umuA}
\ee
where the $i=\theta, \varphi$ components are
\be
u_i=\sum_{l,m} \delta u_{lm} (t,r) E_{ij} \partial^j Y_{lm} (\theta, \varphi)\,.
\ee
The perturbation $\delta u_{lm}$ is a function of $t$ and $r$. 
We expand the action (\ref{action}) up to second order in odd-parity 
perturbations.
In doing so, we can set $m=0$ without loss of generality 
and multiply the action $2\pi$ for the integral 
with respect to $\varphi$ \cite{Kase:2018voo}. 
In the following, we also omit the subscripts ``$lm$'' from 
the variables $Q_{lm}$, $W_{lm}$, and $\delta u_{lm}$ 
for the simplification of notation. 
On using the background Eqs.~(\ref{back1})-(\ref{back3}), 
the resulting second-order action of odd-parity perturbations 
is expressed in the form 
\be
{\cal S}_{\rm odd}=\sum_{l,m}L \int {\rm d}t {\rm d} r\,{\cal L}_{\rm odd}\,,
\ee
where 
\be
L:=l(l+1)\,,
\ee
and 
\begin{widetext}
\ba
{\cal L}_{\rm odd}
&=&
\frac{r^2}{16\pi G_{\ae}} \sqrt{\frac{f}{h}}
\biggl[ C_1\left( \dot{W}-Q'+\frac{2}{r}Q \right)^2
+2\left(C_2\dot{\delta u}+C_3\delta u'+C_4\delta u \right)
\left(\dot{W}-Q'+\frac{2}{r}Q\right)
+C_5\dot{\delta u}^2
+C_6\dot{\delta u}\delta u' \nonumber \\
& &+C_7{\delta u}'^2+(L-2)
\left( C_8W^2+C_9 W\delta u
-a C_9W Q
+C_{10}Q^2+C_{11}Q\delta u \right)
+(LC_{12}+C_{13})\delta u^2\biggr]\,,
\label{oddLag}
\ea
\end{widetext}
with a dot being the derivative with respect to $t$.
The coefficients $C_i$'s in Eq.~(\ref{oddLag}) 
are given in Appendix B. 
Even with the unit-vector constraint $u^{\alpha} u_{\alpha}+1=0$  
in Einstein-Aether theory, the Lagrangian (\ref{oddLag}) is of the same 
form as that derived for generalized Proca theories \cite{Kase:2018voo}, 
with the correspondence of the temporal vector component 
$A_0 \to -af$. The difference appears only for the coefficients $C_i$'s, 
so we can resort to the prescription exploited in Ref.~\cite{Kase:2018voo} 
for the derivation of stability conditions of dynamical perturbations.

Let us consider the gauge transformation $x _{\mu} \to x_{\mu}+\xi_{\mu}$, where
\be
\xi_t=\xi_r=0\,,\qquad 
\xi_i=\sum_{lm}\Lambda(t,r) E_{ij} \partial^j Y_{lm}(\theta,\varphi)\,.
\ee
Then, the perturbations $Q$, $W$, and $\delta u$ 
transform as \cite{Kobayashi,EFT,Kase:2018voo}
\ba
Q &\to& Q+\dot{\Lambda}\,,\label{Qtra}\\
W &\to& W+\Lambda'-\frac{2\Lambda}{r}\,,\label{Wtra}\\
\delta u &\to& \delta u\,.
\ea
Besides the Aether perturbation $\delta u$, we consider 
the following gauge-invariant combination
\be
\chi=\dot{W}-Q'+\frac{2}{r}Q
+\frac{C_2 \dot{\delta u}+C_3 \delta u'+C_4 \delta u}{C_1}\,,
\label{chidef}
\ee
which is associated with the tensor perturbation in the 
odd-parity gravity sector. 
The gauge-invariant perturbation (\ref{chidef}) is introduced to 
combine the first and second contributions to the square brackets 
of Eq.~(\ref{oddLag}). 
We express the Lagrangian (\ref{oddLag}) in the form
\begin{widetext}
\ba
{\cal L}_{\rm odd}
&=&
\frac{r^2}{16\pi G_{\ae}} \sqrt{\frac{f}{h}}
\biggl[ C_1 \biggl\{ 2\chi \biggl( 
\dot{W}-Q'+\frac{2}{r}Q
+\frac{C_2 \dot{\delta u}+C_3 \delta u'+C_4 \delta u}{C_1}
\biggr)-\chi^2 \biggr\} 
-\frac{(C_2 \dot{\delta u}+C_3 \delta u'+C_4 \delta u)^2}{C_1}
+C_5\dot{\delta u}^2\nonumber \\
& &
+C_6\dot{\delta u}\delta u' +C_7{\delta u}'^2+(L-2)
\left( C_8W^2+C_9 W\delta u
-a C_9W Q
+C_{10}Q^2+C_{11}Q\delta u \right)
+(LC_{12}+C_{13})\delta u^2\biggr]\,,
\label{oddLag2}
\ea
\end{widetext}
where $\chi$ is regarded as a Lagrange multiplier independent 
of the fields $W$ and $Q$ in Eq.~(\ref{oddLag2}).
The similar treatment was also performed in the context 
of scalar-tensor theories \cite{DST,Kobayashi,EFT} and 
generalized Proca theories \cite{Kase:2018voo}.

Varying Eq.~(\ref{oddLag2}) with respect to $W$ and $Q$, 
it follows that 
\ba
\hspace{-0.3cm}
&&
2C_1\dot{\chi}-(L-2)\left[2C_8W
+C_9 \left(\delta u-a Q
\right)\right]=0\,,
\label{eqW}\\
\hspace{-0.3cm}
&&
2C_1\chi'+\frac{2rfhC_1'+(8fh+rf'h-rfh')C_1}{rfh}\chi 
\nonumber \\
& &
-(L-2)\left( a C_9 W-2C_{10}Q
-C_{11}\delta u \right)=0\,. 
\label{eqQ}
\ea
These equations can be solved for $W$ and $Q$ to express 
them in terms of $\chi$, $\dot{\chi}$, $\chi'$, and $\delta u$. 
Substituting them into Eq.~(\ref{oddLag2}) and integrating  
it by parts, we obtain the reduced Lagrangian 
\ba
\hspace{-0.7cm}
{\cal L}_{\rm odd}
&=&\frac{r^2}{16 \pi G_{\ae}(L-2)} \sqrt{\frac{f}{h}}
( \dot{\vec{\mathcal{X}}}^{t}{\bm K}\dot{\vec{\mathcal{X}}}
+\dot{\vec{\mathcal{X}}}^{t}{\bm R}\vec{\mathcal{X}}'
\nonumber \\
\hspace{-0.7cm}
& &
+\vec{\mathcal{X}}'^{t}{\bm G}\vec{\mathcal{X}}' 
+\vec{\mathcal{X}}^{t}{\bm M}\vec{\mathcal{X}})\,,
\label{LM2}
\ea
where 
\be
\vec{\mathcal{X}}^{t}=\left( \chi,\delta u \right)\,,
\label{Xt}
\ee
and ${\bm K}, {\bm R}, {\bm G}, {\bm M}$ 
are $2\times2$ symmetric matrices. 
We note that the contributions to Eq.~(\ref{LM2}) of the forms 
$\dot{\vec{\mathcal{X}}}^{t}{\bm T}\vec{\mathcal{X}}$ and 
$\vec{\mathcal{X}}'^{t}{\bm S}\vec{\mathcal{X}}$, 
which appear in generalized Proca 
theories \cite{Kase:2018voo}, vanish in Einstein-Aether theory.
The Lagrangian (\ref{LM2}) can now be used to study the stability of 
dynamical fields $\chi$ and $\delta u$.

The nonvanishing components of ${\bm K}$ are given by 
\be
K_{11}=q_1\,,\qquad 
K_{22}=(L-2)q_2\,,
\label{Kmat}
\ee
where 
\be
q_1:=\frac{4C_1^2 C_{10}}{a^2 C_9^2-4C_8 C_{10}}\,,\qquad 
q_2:=\frac{C_1C_5-C_2^2}{C_1}\,.
\label{q12}
\ee
To avoid the appearance of ghosts, we require that 
\ba
& &
q_1>0\,,
\label{noghost0}\\
& & 
q_2>0\,,
\label{noghost}
\ea
where the former and latter correspond to the no-ghost 
conditions of gravity and vector-field sectors, respectively.

The matrices ${\bm R}$ and ${\bm G}$ have the following 
nonvanishing components
\ba
\hspace{-0.8cm}
& &
R_{11}={\cal R}_{11} q_1,\qquad 
R_{22}=(L-2){\cal R}_{22}\,,\label{Rmat}\\
\hspace{-0.8cm}
&&
G_{11}={\cal G}_{11} q_1\,,\qquad 
G_{22}=(L-2){\cal G}_{22}\,,
\label{Gmat}
\ea
where 
\ba
\hspace{-0.5cm}
& &
{\cal R}_{11}:=-\frac{a C_9}{C_{10}}\,,\qquad
{\cal R}_{22}:=\frac{C_1C_6-2C_2C_3}{C_1}\,,\\
\hspace{-0.5cm}
& &
{\cal G}_{11}:=\frac{C_8}{C_{10}}\,,\qquad
{\cal G}_{22}:=\frac{C_1C_7-C_3^2}{C_1}\,.
\ea

To derive the dispersion relation along the radial direction, we 
assume the solutions of Eq.~(\ref{Xt}) in the form 
$\vec{\mathcal{X}}^{t}=\vec{\mathcal{X}}_0^{t} e^{i( \omega t-kr)}$, 
where $\vec{\mathcal{X}}_0^{t}$ is a constant vector, and 
$\omega$ and $k$ are the constant frequency and wavenumber 
respectively. In the limits $k \to \infty$ and $\omega \to \infty$, 
the existence of nonvanishing solutions of $\vec{\mathcal{X}}^{t}$ 
requires that
${\rm det}(\omega^2{\bm K}-\omega k {\bm R}+k^2 {\bm G})=0$.
Since there are no off-diagonal components in 
${\bm K}$, ${\bm R}$, and ${\bm G}$, it follows that 
\ba
& &
\omega^2-\omega k {\cal R}_{11}+k^2 {\cal G}_{11}=0\,,
\label{dispe1}\\
& &
\omega^2 q_{2}-\omega k {\cal R}_{22}+k^2 {\cal G}_{22}=0\,.
\label{dispe2}
\ea

In terms of the proper time $\tau=\int \sqrt{f}\,{\rm d}t$ and 
the rescaled radial coordinate $r_*=\int {\rm d}r/\sqrt{h}$, 
the propagation speed of perturbations along the radial direction 
is given by $c_r={\rm d}r_*/{\rm d}\tau=\hat{c}_r/\sqrt{fh}$,
where $\hat{c}_r={\rm d}r/{\rm d}t=\omega/k$ 
is the propagation speed in the coordinates $t$ and $r$.
Substituting $\omega=k \sqrt{fh}\,c_r$ into Eqs.~(\ref{dispe1}) 
and (\ref{dispe2}), the solutions to $c_r$ are given, respectively, by 
\ba
c_{r1}&=& \frac{{\cal R}_{11} \pm \sqrt{{\cal F}_1}}{2\sqrt{fh}}
\,,\label{cr1}\\
c_{r2}&=& \frac{{\cal R}_{22} \pm \sqrt{{\cal F}_2}}{2q_2\sqrt{fh}}\,,
\label{cr2}
\ea
where
\ba
\hspace{-0.6cm}
{\cal F}_1 &:=& {\cal R}_{11}^2-4{\cal G}_{11}\,,
\label{F1con}\\
\hspace{-0.6cm}
{\cal F}_2 &:=& {\cal R}_{22}^2-4q_2 {\cal G}_{22}\,.
\label{F2con}
\ea
The speeds $c_{r1}$ and $c_{r2}$ correspond to the radial sound speeds 
associated with the propagation of gravity and vector-field sectors, 
respectively. Depending on the direction of radial propagation, 
the signs of $c_{r1}$ and $c_{r2}$ can be either positive or negative.
As long as $c_{r1}$ and $c_{r2}$ are real, we have 
$c_{r1}^2 \geq 0$ and $c_{r2}^2 \ge 0$. 
Hence the absence of Laplacian instabilities 
along the radial direction requires that 
\ba
& &
{\cal F}_1 \geq 0\,,
\label{F1dcon}\\
& &
{\cal F}_2 \geq 0\,.
\label{F12}
\ea

The propagation speed $c_{\Omega}$ along the angular direction 
can be derived by taking the limits $L \to \infty$ and 
$\omega \to \infty$ in Eq.~(\ref{LM2}). 
In these limits, the dominant contributions to the matrix components 
of ${\bm M}$ are given by 
\be
M_{11}=-LC_1\,,\qquad
M_{22}=L^2 D_1\,,
\label{defD2}
\ee
where 
\be
D_1:=
C_{12}+\frac{C_8C_{11}^2+C_9^2(C_{10}+aC_{11})}
{4C_1^2C_{10}}q_1\,.
\label{D2}
\ee
There are also the matrix components $M_{12}\,(\,=M_{21})$  
proportional to $L$, but they do not affect the angular sound 
speeds derived below. 
Substituting the solution of the form 
$\vec{\mathcal{X}}^{t}=\vec{\mathcal{X}}_0^{t}
e^{i(\omega t-l \theta)}$ into the perturbation equations 
following from (\ref{LM2}), we 
obtain the dispersion relation 
${\rm det}(\omega^2{\bm K}+{\bm M})=0$.  
There are no off-diagonal components of ${\bm K}$ 
and ${\bm M}$, so that 
\ba
& &
\omega^2 q_1-LC_1=0\,,\label{Omere1}\\
& &
\omega^2 q_2+L D_1=0\,.\label{Omere2}
\ea
The angular propagation speed in proper time is given by 
$c_{\Omega}=r {\rm d}\theta/{\rm d}\tau=
\hat{c}_{\Omega}/\sqrt{f}$, where 
$\hat{c}_{\Omega}=r{\rm d}\theta/{\rm d}t$ 
satisfies $\omega^2=\hat{c}_{\Omega}^2l^2/r^2$. 
Taking the limit $L \to \infty$ and substituting the relation 
$\omega^2=c_{\Omega}^2f L/r^2$ into 
Eqs.~(\ref{Omere1}) and (\ref{Omere2}), 
the solutions to $c_{\Omega}^2$ are given, respectively, by 
\ba
& &
c_{\Omega 1}^2=\frac{C_1 r^2}{f q_1}\,,
\label{cOme1si}\\
& &
c_{\Omega 2}^2=-\frac{D_1 r^2}{f q_2}\,.
\ea
To avoid the Laplacian instabilities along the angular direction, 
we require that 
\ba
& &
c_{\Omega 1}^2 \geq 0\,,
\label{cOmecon0}\\
& &
c_{\Omega 2}^2 \geq 0\,,
\label{cOmecon}
\ea
which translate to $C_1/q_1 \geq 0$ and $D_1/q_2 \leq 0$, 
respectively, outside the horizon ($f>0$).

So far, we have considered the stabilities of perturbations 
$\chi$ and $\delta u$ along the radial and angular directions 
by separately taking the limits $k \to \infty$ or $L \to \infty$. 
We will also study the propagation of inclined modes where 
the limits $k \to \infty$ and $L \to \infty$ are taken, with 
the ratio
\be
\xi:= \frac{k}{\sqrt{L}}, 
\ee
being constant. 
In this case, we substitute the solution 
$\vec{\mathcal{X}}^{t}=\vec{\mathcal{X}}_0^{t} 
e^{i( \omega t-kr-l \theta)}$ into the perturbation equations 
following from Eq.~(\ref{LM2}). Then, the dispersion relation 
yields ${\rm det}(\omega^2{\bm K}-\omega k {\bm R}
+k^2 {\bm G}+{\bm M})=0$, so that 
\ba
& &
q_1 \left( \omega^2 -\omega k {\cal R}_{11}+k^2 {\cal G}_{11} 
\right) -LC_1=0\,,
\label{disin1} \\
& &
\omega^2 q_{2}-\omega k {\cal R}_{22}+k^2 {\cal G}_{22}+LD_1=0\,.
\label{disin2}
\ea
Solving Eqs.~(\ref{disin1}) and (\ref{disin2}) for $\omega$ respectively, 
we obtain the dispersion relations for the perturbations $\chi$ and 
$\delta u$, as
\ba
\omega &=& \frac{\sqrt{L}}{2} \left[ {\cal R}_{11} \xi
\pm \sqrt{{\cal F}_1 \xi^2+\frac{4C_1}{q_1}} 
\right]\,,\\
\omega &=& \frac{\sqrt{L}}{2q_2} \left[ {\cal R}_{22} \xi
\pm \sqrt{{\cal F}_2 \xi^2-4q_2 D_1} 
\right]\,,
\ea
where ${\cal F}_1$ and ${\cal F}_2$ are defined by 
Eqs.~(\ref{F1con}) and (\ref{F2con}). 
The absence of Laplacian instabilities for the 
perturbations $\chi$ and $\delta u$ can be ensured 
under the conditions
\ba
& &
{\cal F}_1 \xi^2+\frac{4C_1}{q_1} \geq 0\,,\label{incge}\\
& &
{\cal F}_2 \xi^2-4q_2 {\cal D}_1 \geq 0\,,
\label{incged}
\ea
respectively. 
Let us first consider the stability of the perturbation $\chi$.
In the limit $\xi \to \infty$, there is no Laplacian 
instability for ${\cal F}_1 \geq 0$, 
see Eq.~(\ref{F1dcon}).
In the other limit $\xi \to 0$, the angular propagation speed 
squared is given by Eq.~(\ref{cOme1si}), so the stability 
is ensured for $C_1/q_1 \geq 0$.
Under these two conditions, the inequality (\ref{incge}) 
holds for any arbitrary values of $\xi$. 
On using Eq.~(\ref{incged}), we also find that 
the same property holds for the perturbation $\delta u$, 
i.e., the conditions (\ref{F12}) and (\ref{cOmecon}) are sufficient to 
ensure the Laplacian stability of the inclined mode.
Hence the inclined mode does not give rise to 
additional conditions to those derived for purely radial 
and angular modes.

In summary, for $l \geq 2$, the stabilities of perturbations 
for high radial and angular 
momentum modes are ensured under the conditions 
(\ref{noghost0}), (\ref{noghost}), (\ref{F1dcon}), (\ref{F12}), 
(\ref{cOmecon0}) and (\ref{cOmecon}). 
We caution that these conditions are derived in the large 
$k$ or (and) $l$ limits, so they are not sufficient to guarantee 
all the stabilities for finite values of $k$ and $l$. 
Due to the complexity of matrix components of ${\bm M}$, 
we do not consider the stability of perturbations for such 
an intermediate range of $k$ and $l$.

In addition, instabilities might arise when we consider the spectrum 
of $\vec{\mathcal{X}}^{t}$, by solving the corresponding differential equations for
$\chi$ and $\delta u$ with boundary conditions. However, such studies 
are out of the scope of the current
paper, and we wish to return to this important issue in another occasion. 

\subsection{$l=1$}

Since the metric components $h_{ij}$ vanish identically for
the dipole perturbation ($l=1$), there is a gauge degree of freedom 
to be fixed. In this case, we choose the gauge $W=0$. 
{}From Eq.~(\ref{Wtra}), the gauge-transformation scalar $\Lambda(t, r)$ 
is constrained to be 
\be
\Lambda=-r^2 \int {\rm d}\tilde{r}\,\frac{W(t,\tilde{r})}{\tilde{r}^2}
+r^2 {\cal C}(t)\,,
\label{Lambdain}
\ee
where ${\cal C}(t)$ is a function of $t$.
The Lagrangian (\ref{oddLag}) has been derived for $l \geq 2$ 
with a nonvanishing $W$, but it is also valid for $l=1$ by 
setting $W=0$ for the above gauge choice.
An alternative procedure to be taken for $l=1$ is that  
we literally exploit the Lagrangian (\ref{oddLag}), 
vary it with respect to 
$W$ and $Q$,  and set $W=0$ at the end. 
This process leads to 
\be
\dot{\cal E}=0\,,\qquad 
\left(r^2{\cal E}\right)'=0\,,
\label{Eeq}
\ee
where 
\be
{\cal E}:= r^2 \sqrt{\frac{f}{h}} 
\left[ C_1\left(Q'-\frac{2}{r}Q \right)-\left( C_2\dot{\delta u}
+C_3\delta u'+C_4\delta u\right) \right].
\label{calE}
\ee
{}From Eq.~(\ref{Eeq}), we obtain the integrated solution 
\be
{\cal E}=\frac{{\cal E}_0}{r^2}\,,
\label{Eint}
\ee
where ${\cal E}_0$ is a constant. 
On using Eq.~(\ref{calE}), the perturbation $Q$ can be 
expressed as 
\ba
Q &=&
r^2\int {\rm d} \tilde{r} \frac{1}{C_1 \tilde{r}^2} 
\left( \frac{{\cal E}_0}{\tilde{r}^4}\sqrt{\frac{h}{f}}
+C_2\dot{\delta u}+C_3\delta u'+C_4 \delta u \right)
\nonumber \\
& &
+r^2{\cal C}_2(t) \,,
\label{Qin}
\ea
where ${\cal C}_2$ is a function of $t$.
From Eqs.~(\ref{Qtra}) and (\ref{Qin}), 
the residual gauge mode ${\cal C}(t)$ in Eq.~(\ref{Lambdain}) 
can be removed by setting
\be
{\cal C}(t)=\int {\rm d}\tilde{t}\,{\cal C}_2(\tilde{t})\,.
\ee
On using Eq.~(\ref{Eint}) with Eq.~(\ref{calE}), we can eliminate 
the terms containing $-Q'+2Q/r$ in Eq.~(\ref{oddLag}). 
This process leads to the reduced Lagrangian
\begin{widetext}
\ba
{\cal L}_{\rm odd}=
&&
\frac{r^2}{16\pi G_{\ae}} \sqrt{\frac{f}{h}} \Bigg[ 
\frac{C_1C_5-C_2^2}{C_1}\dot{\delta u}^2
+\left(C_6-\frac{2C_2C_3}{C_1} \right)\dot{\delta u} 
\delta u' +\left(C_7-\frac{C_3^2}{C_1} \right)\delta u'^2
\notag\\
&&
\hspace{2cm}
-\frac{2C_3C_4}{C_1}\delta u' \delta u
+\left(2C_{12}+C_{13}-\frac{C_4^2}{C_1}\right)
\delta u^2+\frac{h{\cal E}_0^2}{C_1r^8f}
\Bigg]\,.
\label{Ldipole}
\ea
\end{widetext}
This shows that only the Aether perturbation $\delta u$ 
propagates. {}From the coefficient of $\dot{\delta u}^2$ in Eq.~(\ref{Ldipole}), 
we find that the ghost is absent for $(C_1 C_5-C_2^2)/C_1>0$, 
which is equivalent to the condition $q_2>0$ derived for $l \geq 2$.
{}From the first three terms in Eq.~(\ref{Ldipole}), we can also 
show that the radial propagation speed is equivalent to 
$c_{r2}$ given by Eq.~(\ref{cr2}). 
This is analogous to the result found for generalized 
Proca theories \cite{Kase:2018voo}.
In summary, for $l=1$, there are neither ghost nor Laplacian instabilities 
under the conditions $q_2>0$ and ${\cal F}_2 \geq 0$. 

\section{Stability conditions in Einstein-Aether theories}
\label{stabilitysec}

To study the odd-parity stabilities of BHs, 
we use the explicit forms of $C_i$'s given in Appendix B
together with the value of $b$ constrained by Eq.~(\ref{bso}). 
Then, the no-ghost conditions (\ref{noghost0}) and 
(\ref{noghost}) translate to 
\be
q_1 = \frac{h(1-c_{13})(1-c_{13}a^2 f)}{2f^2}>0\,,
\label{stacon1}
\ee
\be
q_2 = \frac{c_1+c_4 a^2 f}{r^2 f}
-\frac{c_{13}^2 (a^2f-1)}{2r^2 f(1-c_{13})}>0\,.
\label{stacon2}
\ee
The conditions (\ref{F1dcon}) and (\ref{F12}), which ensure the absence of 
Laplacian instabilities along the radial direction, 
are given by 
\ba
{\cal F}_1 &=& \frac{4fh(1-c_{13})}{(1-c_{13}a^2 f)^2} \geq 0\,,
\label{stacon3}\\
\hspace{-1cm}
{\cal F}_2 &=& \frac{2h c_{14}
[2c_1-c_{13}(2c_1-c_{13})]}{r^4 f (1-c_{13})} \geq 0\,.
\label{stacon4}
\ea

The conditions (\ref{cOmecon0}) and (\ref{cOmecon}) for the absence of 
Laplacian instabilities along the angular direction 
translate to 
\ba
\hspace{-0.7cm}
c_{\Omega 1}^2 &=& \frac{1}{1-c_{13}a^2 f} \geq 0\,,
\label{stacon5}\\
\hspace{-0.7cm}
c_{\Omega 2}^2 &=& \frac{2c_1-c_{13} (2c_1-c_{13})}
{2(1-c_{13})(c_1+c_4 a^2f)-c_{13}^2 (a^2f-1)} \geq 0.
\label{stacon6}
\ea
For the dipole ($l=1$), only the two conditions 
(\ref{stacon2}) and (\ref{stacon4}) need to be satisfied 
for the Aether perturbation $\delta u$.

On the Minkowski background characterized by 
the metric components $f=h=1$, the Aether field is
given by $u^{\mu}=(+1,0,0,0)$ and hence $a=1$ and $b=0$. 
Then, in Minkowski spacetime, the stability conditions 
(\ref{stacon1})-(\ref{stacon6}) reduce, respectively, to 
\ba
\hspace{-0.8cm}
(q_1)_{\rm Min} &=&\frac{(1-c_{13})^2}{2}>0\,,\\ 
\hspace{-0.8cm}
(q_2)_{\rm Min} &=&\frac{c_{14}}{r^2}>0\,,\label{q2Min}\\
\hspace{-0.8cm}
({\cal F}_1)_{\rm Min} &=& \frac{4}{1-c_{13}} \geq 0\,,\\
\hspace{-0.8cm}
({\cal F}_2)_{\rm Min} &=&
\frac{2 c_{14}
[2c_1-c_{13}(2c_1-c_{13})]}{r^4 (1-c_{13})}  \geq 0\,,\\
\hspace{-0.8cm}
(c_{\Omega 1}^2)_{\rm Min} &=& \frac{1}{1-c_{13}} \geq 0\,,\\
\hspace{-0.8cm}
(c_{\Omega 2}^2)_{\rm Min} &=& \frac{2c_1-c_{13} (2c_1-c_{13})}
{2c_{14}(1-c_{13})} \geq 0\,,
\ea
which are satisfied for 
\ba
& &
c_{13}<1\,,\\
& &
c_{14} >0\,,\label{c14}\\
& &
2c_1-c_{13} (2c_1-c_{13}) \geq 0\,.
\label{c113}
\ea
These conditions coincide with those
derived  in Refs.~\cite{Jacobson:2004ts,Oost:2018tcv} by expanding 
the action (\ref{action}) up to quadratic 
order in tensor and vector perturbations on the Minkowski background. 
On using the fact that the coefficients $C_9$, $C_2$, and $C_6$ vanish 
in Eqs.~(\ref{cr1}) and (\ref{cr2}), the radial sound speed squares 
in Minkowski spacetime are given by 
\ba
(c_{r 1}^2)_{\rm Min} &=& \frac{1}{1-c_{13}}\,, \\
(c_{r 2}^2)_{\rm Min} &=& \frac{2c_1-c_{13} (2c_1-c_{13})}
{2c_{14}(1-c_{13})}\,.
\ea
Then we have that 
$(c_{r 1}^2)_{\rm Min}=(c_{\Omega 1}^2)_{\rm Min}$ 
and $(c_{r 2}^2)_{\rm Min}=(c_{\Omega 2}^2)_{\rm Min}$, 
while this equality does not generally hold on 
the curved background (\ref{metric}).

As shown in Refs.~\cite{Jacobson:2004ts,Oost:2018tcv}, $(c_{r 1}^2)_{\rm Min}$ and 
$(c_{r 2}^2)_{\rm Min}$ correspond to the propagation speed 
squares of tensor and vector perturbations on the Minkowski background, 
respectively, see Eqs.~(\ref{cTm}) and (\ref{cVm}).
From the gravitational-wave event GW170817 \cite{GW170817} together 
with its electromagnetic counterpart \cite{Goldstein}, the speed squared 
of tensor perturbations is in the range 
$-3 \times 10^{-15}<(c_{r 1})_{\rm Min}-1<7 \times 10^{-16}$, 
so the coupling $c_{13}$ is constrained to be 
\be
|c_{13}| \lesssim 10^{-15}\,.
\label{gwcon}
\ee
We note that there are also stability conditions in the Minkowski spacetime  
arising from scalar perturbations \cite{Jacobson:2004ts,Oost:2018tcv}.
They can be derived by considering even-parity perturbations 
on the curved background (\ref{metric}) and taking the 
Minkowski limit $f \to 1$, $h \to 1$, and $a \to 1$.

In this paper, we will not carry out the analysis of even-parity perturbations, 
but we will show in Sec.~\ref{BHstasec} that the stability analysis 
based on odd-parity perturbations alone is sufficiently powerful 
to exclude some BH solutions in Einstein-Aether theory.

\section{Stability of Einstein-Aether black holes}
\label{BHstasec}

Let us consider the stability of spherically symmetric and static 
BH solutions known in the literature.  
Performing the transformation 
\be
{\rm d}v={\rm d}t+\frac{{\rm d}r}{\sqrt{fh}}\,,
\label{vttra}
\ee
the line element (\ref{metric}) is transformed to 
the Eddington-Finkelstein coordinate of the form 
\be
{\rm d}s^{2} =-f(r) {\rm d}v^{2} +2B(r){\rm d}v{\rm d}r + 
r^{2} \left( {\rm d}\theta^{2}+\sin^{2}\theta\, 
{\rm d} \varphi^{2} \right)\,,
\label{metric2}
\ee
where 
\be
B(r)=\sqrt{\frac{f}{h}}\,.
\label{Br}
\ee
{}From Eq.~(\ref{umuA}), the nonvanishing components 
of the background Aether field $u_{\mu}$ are given by 
$u_t=-af$ and $u_r=b/h$, where $b$ is constrained as 
Eq.~(\ref{bso}).
On using Eq.~(\ref{vttra}), we have 
\be
u_t {\rm d}t+u_r {\rm d}r=u_v {\rm d}v+\tilde{u}_r {\rm d}r\,,
\ee
where 
\be
u_v=-af\,,\qquad 
\tilde{u}_r=a \sqrt{\frac{f}{h}}+\frac{b}{h}\,.
\ee
Since $g^{vv}=0$, $g^{vr}=g^{rv}=\sqrt{h/f}$, and 
$g^{rr}=h$, the nonvanishing components of $u^{\mu}$ 
in the $(v,r)$ coordinate are given by 
\be
u^v=a+\frac{b}{\sqrt{fh}}\,,\qquad 
\tilde{u}^r=b\,.
\ee
In the notation of Ref.~\cite{Zhang:2020too}, the variable 
$A$ is used for $u^v$, in which case we have 
\be
u^v=A=a+\frac{b}{\sqrt{fh}}\,,\qquad 
\tilde{u}^r=b=\frac{fA^2-1}{2AB}\,.
\label{Au}
\ee

Since we would like to consider the case in which 
the gravitational-wave bound (\ref{gwcon}) is satisfied, 
we will focus on the BH solutions satisfying 
the conditions
\be
c_{13}=0\,,
\label{c13con}
\ee
in the following analysis.

\subsection{Stealth Schwarzschild solution}
\label{stesec}

We first consider the coupling constants satisfying 
\be
c_{14} =0\,.
\label{c14con}
\ee
For this choice we have 
$(q_2)_{\rm Min}=0$ from Eq.~(\ref{q2Min}), so 
there is a strong coupling problem on the Minkowski background. 
In curved spacetime the stability conditions 
are different from those on the Minkowski background, 
so we will study whether BH solutions satisfying 
the condition (\ref{c14con}) are stable or not.

The background Eqs.~(\ref{back1})-(\ref{back3}) 
admit the existence of an exact stealth BH 
solution characterized by
\ba
& &
f=h=1-\frac{r_s}{r}\,,\label{fhana} \\
& & 
a=\frac{\sqrt{4r^3 (r-r_s)+w_2^2}}{2r(r-r_s)}\,,\qquad
b=\epsilon \frac{w_2}{2r^2}\,,
\label{aana}
\ea
where $r_s$ is the Schrawarzschild radius, $\epsilon= \pm 1$, 
and $w_2$ is a positive constant. 
{}From Eqs.~(\ref{Br}) and (\ref{Au}), we have 
\be
A=\frac{\epsilon w_2+\sqrt{4r^3 (r-r_s)+w_2^2}}
{2r (r-r_s)}\,,\qquad 
B=1\,,\label{ABana}
\ee
so that $A \to 1$ as $r \to \infty$.
For $\epsilon=+1$, the temporal vector component 
diverges as $u^v=A \propto (r-r_s)^{-1}$ 
around $r=r_s$. 
In this case, the quantity $J:=A^2 f$ is in the range $J>1$ 
outside the horizon and it exhibits the divergence
$J \to (w_2^2/r_s^3)(r-r_s)^{-1}$ for $r \to r_s$.
On the other hand, for $\epsilon=-1$, 
the expansion of $A$ around $r=r_s$ gives 
\be
A=\frac{r_s^2}{w_2}-\frac{r_s (r_s^4-2w_2^2)}{w_2^3}
(r-r_s)+{\cal O} ((r-r_s)^2)\,,
\label{Are}
\ee
and hence $A$ is finite at $r=r_s$. 
In this case, $J<1$ outside the horizon and $J=0$ 
at $r=r_s$. For this latter branch ($b<0$), the above exact 
BH solution with $w_2=3\sqrt{3}r_s^2/8$ gives rise to 
a universal horizon at $r=3r_s/4$ \cite{Zhang:2020too}.

{}From Eq.~(\ref{aana}), the temporal vector component 
$u^t$ in the $(t, r)$ coordinate has the divergent behavior 
$u^t=a \propto (r-r_s)^{-1}$ as $r \to r_s$, irrespective of 
the signs of $b$. Defining the quantity 
\be
j:=a^2 f\,,
\label{jdef}
\ee
we have
\be
j=\frac{w_2^2}{4r^3 (r-r_s)}+1\,,
\ee
and hence $j>1$ outside the horizon.
There is the divergence $j \to \infty$ as $r \to +r_s$, with 
the asymptotic behavior $j \to 1$ at spatial infinity.

{}From Eqs.~(\ref{stacon1})-(\ref{stacon6}), the quantities 
associated with the stability conditions are given by 
\ba
& &
q_1=\frac{1}{2f}\,,\qquad q_2=-\frac{c_1 (j-1)}{r^2 f}\,,\\
& &
{\cal F}_1=4f^2\,,\qquad {\cal F}_2=0\,,\\
& &
c_{\Omega 1}^2=1\,,\qquad c_{\Omega 2}^2=-\frac{1}{j-1}\,.
\ea
The conditions $q_1>0$, ${\cal F}_1 \geq 0$, 
${\cal F}_2 \geq 0$, and $c_{\Omega 1}^2 \geq 0$ are satisfied 
for $r>r_s$. Since $j>1$ outside the horizon, we have $c_{\Omega 2}^2<0$ 
and hence there is a Laplacian instability along the angular direction. 
In particular, as $r$ increases from $r_s$ to spatial infinity, 
$c_{\Omega 2}^2$ changes from $-0$ to $-\infty$.
For $c_1>0$, we have $q_2<0$ outside the horizon, so the ghost 
instability is also present.
{}From Eqs.~(\ref{cr1}) and (\ref{cr2}), the radial propagation speed 
squares are given by 
\be
c_{r1}^2=1\,,\qquad 
c_{r2}^2=\frac{j}{j-1}\,.
\ee
Since both $c_{r1}^2$ and $c_{r2}^2$ are positive outside the horizon, 
the Laplacian instabilities are absent along the radial direction. 

In summary, the stealth Schwarzschild solution with the vector-field
profile (\ref{aana}) is unstable due to the Laplacian instability associated 
with the negative propagation speed squared $c_{\Omega2}^2$
outside the horizon. 
In addition, for $c_1 > 0$, the ghost instability for
the Aether perturbation also exists.
It is interesting to note that stealth Schwarzschild solutions present in 
the context of generalized Proca 
theories \cite{Chagoya:2016aar,Heisenberg:2017xda,Heisenberg:2017hwb} 
are also unstable against 
odd-parity perturbations \cite{Kase:2018voo}.

\subsection{BH solutions with $c_{14} \neq 0$}
\label{BHc1nc4}

We proceed to study the stability of BH solutions 
for the couplings 
\be
c_{14} \neq 0\,.
\ee
Then, the quantities in Eqs.~(\ref{stacon1})-(\ref{stacon6}) 
reduce to 
\ba
& &
q_1=\frac{h}{2f^2}\,,\qquad q_2=\frac{c_1+c_4 j}{r^2 f}\,,
\label{q1ex}\\
& &
{\cal F}_1=4fh\,,\qquad 
{\cal F}_2=\frac{4h c_1 c_{14}}{r^4 f}\,,
\label{F1ex}\\
& &
c_{\Omega 1}^2=1\,,\qquad 
c_{\Omega 2}^2=\frac{c_1}{c_1+c_4 j}\,,
\label{cOex}
\ea
where $j$ is defined by Eq.~(\ref{jdef}).
Since $f>0$ and $h>0$ outside the horizon, 
the conditions $q_1>0$, ${\cal F}_1 \geq 0$, and 
$c_{\Omega 1}^2 \geq 0$ are satisfied.
The condition ${\cal F}_2 \geq 0$ translates to 
\be
c_1 c_{14} \geq 0\,.
\label{c114}
\ee

Taking the asymptotically flat (Minkowski) limit 
$a \to 1$, $f \to 1$, and $h \to 1$ 
in Eqs.~(\ref{q1ex}) and (\ref{cOex}), 
it follows that $q_2 \to c_{14}/r^2$ and 
$c_{\Omega 2}^2 \to c_1/c_{14}$.
Then, the stability conditions $q_2>0$ and 
$c_{\Omega 2}^2 \geq 0$ translate to 
\be
c_{14} >0\,,\qquad c_1 \geq 0\,,
\label{c14c}
\ee
which are compatible with Eq.~(\ref{c114}). 
In the Minkowski limit  the propagation speed 
squares of the transverse vector mode along both 
radial and angular directions are 
$(c_{r2}^2)_{\rm Min}=(c_{\Omega 2}^2)_{\rm Min}
=c_1/c_{14}$, so the propagation is 
subluminal (or superluminal) for $c_4>0$ (or for $c_4<0$). 
In Einstein-Aether theory, the gravitational Cerenkov radiation 
can occur for the subluminal propagation of transverse vector mode. 
For an interaction between a fermion and a graviton studied 
in Ref.~\cite{Elliott:2005va}, the emission rate $\Gamma$ from a fermion 
for the transverse vector mode is proportional to 
$c_{13}^2 [1-(c_{r2}^2)_{\rm Min}]$ \cite{Elliott:2005va}, 
so that $\Gamma=0$ for $c_{13}=0$. 
When $(c_{r2}^2)_{\rm Min}<1$, however, there may be a possibility that other higher-order interactions give rise to the gravitational Cerenkov radiation even for $c_{13}=0$. 
In the superluminal range realized by the coupling $c_4<0$, there is no 
constraint arising from the gravitational Cerenkov radiation.

To discuss the stability of BH solutions around the horizon, 
we search for background solutions where the temporal vector component 
$u^v=A$ in the ($v, r$) coordinate is regular at $r=r_s$ like Eq.~(\ref{Are}).
In doing so, we expand $A$, $f$, $h$ around $r=r_s$ in the forms
\ba
\hspace{-0.7cm}
A &=& A_0+A_1 (r-r_s)+A_2 (r-r_s)^2+\cdots\,,\label{Aite}\\
\hspace{-0.7cm}
f &=& f_1(r-r_s)+f_2 (r-r_s)^2+f_3 (r-r_s)^3\cdots\,,
\label{fite}\\
\hspace{-0.7cm}
h &=& h_1(r-r_s)+h_2 (r-r_s)^2+h_3 (r-r_s)^3\cdots\,,
\label{hite}
\ea
where $A_i$, $f_i$, $h_i$ are constants. 
On using Eqs.~(\ref{Br}) and (\ref{Au}), there is the following relation 
\be
a=\frac{A^2 f+1}{2Af}\,.
\ee
Then, we can express Eqs.~(\ref{back1})-(\ref{back3}) as 
the differential equations for $A$, $f$, $h$, instead of 
those for $a$, $f$, $h$. 
Substituting Eqs.~(\ref{Aite})-(\ref{hite}) into such differential equations, 
we find that there are solutions where the coefficients 
$A_{1,2,3,\cdots}$ and metric components are related to 
the constant $A_0$. 
For the special case with $c_{14}=0$, we confirmed that the iterative 
solutions derived by this prescription coincide with those obtained by 
expanding Eqs.~(\ref{fhana}) and (\ref{ABana}) around $r=r_s$.

{}From Eqs.~(\ref{Aite})-(\ref{hite}), the temporal 
metric component $a$ in the ($t,r$) coordinate and the quantity 
$j=a^2f$ have the following dependence around the horizon:
\ba
a &=& \frac{1}{2A_0 f_1} \left( r-r_s \right)^{-1}
+{\cal O} ((r-r_s)^0)\,,\\
j &=& \frac{1}{4A_0^2 f_1} \left( r-r_s \right)^{-1}
+{\cal O} ((r-r_s)^0)\,.
\ea
Since $j$ diverges at $r=r_s$, the quantities $q_2$ and 
$c_{\Omega 2}^2$ around the horizon can be estimated as
\ba
q_2 &=& \frac{c_4}{4A_0^2 r_s^2 f_1^2} \left( r-r_s \right)^{-2}
+{\cal O} ((r-r_s)^{-1})\,,\label{q2ho} \\
c_{\Omega 2}^2 &=& \frac{4c_1 A_0^2 f_1}{c_4}
\left( r-r_s \right)+{\cal O} ((r-r_s)^{2})\,.
\ea
{}From Eq.~(\ref{q2ho}) the ghost is absent for 
\be
c_4>0\,.
\label{c4p}
\ee
Provided that $c_1 \geq 0$, we also have $c_{\Omega 2}^2 \geq 0$ 
around $r=r_s$.
Indeed, for $c_1 \geq 0$ and $c_4>0$, the two conditions $q_2>0$ 
and $c_{\Omega 2}^2 \geq 0$ 
hold throughout the horizon exterior, since $j$ is positive.
We note that the odd-parity stability about the Minkowski spacetime, 
which is satisfied under the conditions (\ref{c14c}), does not necessarily 
require that $c_4>0$ (unless the superluminality of 
$c_{\Omega 2}^2$ is imposed). 
For BHs  the term $c_4 j$ in Eq.~(\ref{q1ex}) dominates over $c_1$ 
around the horizon, so the positivity of $q_2$ demands that 
$c_4>0$. In other words, the inequality (\ref{c4p}) is a new stability 
condition derived by the analysis on the curved background.

The BH solution with $c_4<0$ and $c_1 > 0$ is plagued by the ghost instability 
as well as the Laplacian instability around the horizon. 
If we restrict the superluminal propagation of transverse vector mode, 
we have $c_4<0$ and hence the BH solution 
in this case is unstable.

So far, we have performed the expansion of Taylor series 
of $A$ as Eq.~(\ref{Aite}) with a finite value of $A$ at $r=r_s$. 
Suppose that there is a solution of $A$ diverging at $r=r_s$
in the form 
\be
A=\frac{A_0}{(r-r_s)^p}\,,
\label{Asin}
\ee
where $A_0$ and $p$ are constants. 
Here we are considering positive values of $p$, 
but we also include the case $p=0$ in the analysis below.
Analogous to the discussion in scalar-tensor theories \cite{Hui:2012qt}, 
we consider the scalar product $J^{\mu \nu}J_{\mu \nu}$ for 
the current tensor ${J^{\mu}}_{\alpha}$ defined in Eq.~(\ref{Jmu}) 
and impose the regularity of $J^{\mu \nu}J_{\mu \nu}$ at $r=r_s$. 
On using Eq.~(\ref{Asin}) and regular expansions of $f$ and $h$ 
as those in Eqs.~(\ref{fite}) and (\ref{hite}), 
the scalar product $J^{\mu \nu}J_{\mu \nu}$ diverges at $r=r_s$ 
apart from the special cases $p=0$ and $p=1$. 
For $p$ close to be 0 or 1, there is the power-law
dependence $J^{\mu \nu}J_{\mu \nu} \propto 1/(r-r_s)^q$ with 
$q$ close to 2. The powers $p=0$ and $p=1$ are the special cases 
in which $J^{\mu \nu}J_{\mu \nu}$ is regular at $r=r_s$. For $p>1$, the scalar product 
diverges as $J^{\mu \nu}J_{\mu \nu} \propto 1/(r-r_s)^{2p}$ at $r=r_s$.

The expansion of $A$ performed in Eq.~(\ref{Aite}) corresponds 
to the power $p=0$, in which case the regularity of $J^{\mu \nu}J_{\mu \nu}$ 
is ensured at the horizon. 
For $p=1$, expanding the quantities $q_2$ and $c_{\Omega 2}^2$
around $r=r_s$ gives 
\ba
q_2 &=& \frac{c_4 A_0^2}{4 r_s^2} \left( r-r_s \right)^{-2}
+{\cal O} ((r-r_s)^{-1})\,,\label{q2ho2} \\
c_{\Omega 2}^2 &=& \frac{4c_1}{c_4 A_0^2 f_1}
\left( r-r_s \right)+{\cal O} ((r-r_s)^{2})\,.
\ea
Under the superluminal condition $c_4<0$, 
there is the ghost instability ($q_2<0$) as well as 
the Laplacian instability 
($c_{\Omega 2}^2<0$) for $c_1>0$. 
As in the case of $p=0$, we require the conditions 
(\ref{c14c}) and (\ref{c4p}) to ensure the odd-parity 
stability of BHs, but in this case the propagation of 
transverse vector mode is subluminal.

\subsection{BH solutions with $c_4=0$}
\label{BHc4=0}

Let us finally discuss the stability of BH solutions for 
the coupling 
\be
c_4=0\,.
\ee
In this case, the quantities $q_1$, ${\cal F}_1$, and $c_{\Omega 1}^2$ are 
the same as those given in Eqs.~(\ref{q1ex}), (\ref{F1ex}), and (\ref{cOex}), 
which are all positive outside the horizon. 
For $c_1 \neq 0$, the other quantities are given by 
\be
q_2=\frac{c_1}{r^2f}\,,\qquad
{\cal F}_2=\frac{4h c_1^2}{r^4 f}\,,\qquad 
c_{\Omega 2}^2=1\,.
\ee
Provided that $c_1>0$, the ghost is absent. 

When $c_1=0$, both the denominator and numerator of $c_{\Omega 2}^2$
in Eq.~(\ref{cOex}) vanish. 
This reflects the fact that, for $c_1=0$, the vector perturbation 
does not propagate as in the case of GR. 
The coupling constant $c_2$ does not appear in any of the 
stability conditions obtained in Sec.~\ref{stabilitysec}, so the case $c_1=0$ 
can be regarded as the GR limit for the couplings 
under consideration now
(i.e., $c_1=0$, $c_3=0$, and $c_4=0$). 
In this case, we only need to consider the stability conditions 
$q_1>0$, ${\cal F}_1 \geq 0$, and $c_{\Omega 1}^2 \geq 0$ 
in the odd-parity sector, 
all of which are trivially satisfied outside the horizon.

In summary, for $c_4=0$, the stability of BHs against 
odd-parity perturbations with large values of $k$ and $l$ 
is ensured for 
\be
c_1 \geq 0\,.
\ee
There are numerically obtained BH solutions consistent 
with this range of couplings \cite{Eling:2006ec,Zhang:2020too}.

\section{Conclusions}
\label{consec}

In this paper, we studied the stability of spherically symmetric 
and static BHs against odd-parity perturbations in Einstein-Aether 
theory. On the background (\ref{metric}), the presence of a unit vector 
constraint (\ref{be1}) gives the relation (\ref{bso}) between the temporal 
and radial components of the Aether field. 
At the background level, there are three independent 
Eqs.~(\ref{back1})-(\ref{back3}) to be solved for $a$ and 
the metric components $f$ and $h$.

In Sec.~\ref{oddparitysec}, we derived the second-order action of 
odd-parity perturbations by using the expansion in terms of the 
spherical harmonics $Y_{lm} (\theta, \varphi)$. Choosing the 
Regge-Wheeler gauge for $l \ge 2$, we obtained the second-order 
Lagrangian of the form (\ref{oddLag}) and identified $\chi$ and $\delta u$ 
as the two dynamical perturbations associated with the gravity sector 
and the Aether field, respectively. 
After the integration by parts, the Lagrangian of these dynamical 
fields is given by Eq.~(\ref{LM2}). 
We showed that there are neither ghost nor Laplacian instabilities 
under the conditions (\ref{noghost0}), (\ref{noghost}), (\ref{F1dcon}), 
(\ref{F12}), (\ref{cOmecon0}) and (\ref{cOmecon}) for large 
values of $k$ and $l$.
For the dipole ($l=1$) the propagating degree of freedom is
the Aether perturbation $\delta u$ alone, 
which does not give additional constraints to those
derived for $l \geq 2$.

Using the explicit forms of coefficients $C_i$'s given in Appendix B,  
the stability conditions in Einstein-Aether theory 
reduce to Eqs.~(\ref{stacon1})-(\ref{stacon6}).
In the limit of the Minkowski spacetime, we also showed 
in Sec.~\ref{stabilitysec} that the propagation 
speeds along both radial and angular directions coincide with those of
tensor and vector perturbations already 
derived in the literature.  
The combination of coupling constants $c_{13}=c_1+c_3$ is 
tightly constrained to be $|c_{13}| \lesssim 10^{-15}$ from 
the GW170817 event together with 170817A. 

In Sec.~\ref{BHstasec}, the odd-parity stabilities of BHs
in Einstein-Aether theory were studied for 
the couplings satisfying 
\be
c_{13}=0\,.
\label{CD2a}
\ee
This choice is consistent with the constraint (\ref{gwcon}) 
on the speed of tensor perturbations in the range 
$-3 \times 10^{-15}<(c_{r 1})_{\rm Min}-1<7 \times 10^{-16}$ 
obtained from the gravitational-wave event GW170817 \cite{GW170817} 
and its electromagnetic counterpart \cite{Goldstein}.
In doing so, we used the relations of metric and vector-field 
components between the two different coordinates 
(\ref{metric}) and (\ref{metric2}). 
In Sec.~\ref{stesec}, we considered the exact Schwarzschild solution
present for $c_{14}=0$ and found that there is a Laplacian 
instability along the angular direction throughout the horizon exterior. 
Moreover, for $c_1 >0$, the ghost instability also exists 
for the Aether perturbation. 
In Sec.~\ref{BHc1nc4}, we discussed the BH solutions 
for $c_{14} \neq 0$ and showed that their stabilities 
require the conditions (\ref{c14c}) and $c_4>0$. 
In this case, the propagation of the vector perturbation
is subluminal in the asymptotically flat regime, so there is 
a possibility for the gravitational Cerenkov radiation to occur. 
In other words, the superluminal propagation of transverse 
vector mode occurring for $c_4<0$, under which the gravitational Cerenkov 
radiation is avoided, is incompatible with the BH stability conditions.
In Sec.~\ref{BHc4=0}, we showed that the BH solutions 
with
\be
c_4=0,\qquad c_1 \geq 0\,,
\lb{CD2b}
\ee
are stable against odd-parity perturbations 
for high radial and angular momentum modes.
Clearly, if we demand the odd-parity stability of BHs,  
the viable region of the parameter space of
Eqs.~(\ref{CD1a})-(\ref{CD1d}), obtained recently in Ref.~\cite{Oost:2018tcv}, 
is reduced further.

It is interesting to note that the instability of perturbations mainly 
happens in the Aether field, represented by $\delta u$, while 
the metric part, represented by $\chi$, behaves well for the 
coupling with $c_{13}=0$.

The Lagrangian (\ref{oddLag}) of odd-parity perturbations can be applied 
to the computation of quasi-normal modes of BHs. 
Moreover, the analysis of even-parity perturbations will provide us 
additional stability conditions of BHs to those derived in this paper.
We leave these issues, together with the analyses of their 
corresponding quasi-normal mode spectra, for future separate publications.

\section*{Acknowledgements}

We thank Ted Jacobson and Shinji Mukohyama for useful discussions 
and comments.
ST is supported by the Grant-in-Aid for Scientific Research 
Fund of the JSPS No.~19K03854. 
This work is also partially supported by the National Key Research 
and Development Program of China under  the Grant No.~2020YFC2201503, 
and the National Natural Science Foundation of China under 
the grant No.~11975203. 

\section*{Appendix A: Qnatities in background equations}
\renewcommand{\theequation}{A.\arabic{equation}} \setcounter{equation}{0}

On the spherically symmetric and static background (\ref{metric}), 
the Lagrange multiplier (\ref{Lagm}) is given by 
\begin{widetext}
\ba
\hspace{-0.5cm}
\lambda & = &
\bigg\{f r^2 \left[a^2 \left(6 c_1+3 c_2+2 c_{13}-8 c_{14}\right) h \left(f'\right)^2+c_2 \left(2 h f''+f' h'\right)\right] \nn\\
\hspace{-0.5cm}
&& +2 a^2 f^4 \left[-2 a^2 \left(2 \left(c_2+c_{13}\right) h-c_2 r h'\right)+4 \left(c_{14}-c_1\right) h r^2 \left(a'\right)^2+a \left(-c_1+c_2+c_{13}\right) r \left(2 h r a''+a' \left(r h'+4 h\right)\right)\right] \nn\\
\hspace{-0.5cm}
&& -f^2 \Big[r h' \left(a^2 \left(-2 c_1+3 c_2+2 c_{13}\right) r f'-4 c_2\right) \nn\\
\hspace{-0.5cm}
&& +2 h \big(a^4 \left(3 c_1+c_2+c_{13}-4 c_{14}\right) r^2 \left(f'\right)^2+a^2 r \left(\left(-2 c_1+3 c_2+2 c_{13}\right) r f''+2 \left(-2 c_1+c_2+2 c_{13}\right) f'\right)\nn\\
\hspace{-0.5cm}
&& -a \left(11 c_1-5 c_2-5 c_{13}-8 c_{14}\right) r^2 a' f'+4 \left(c_2+c_{13}\right)\big)\Big] 
-\left(2 c_2+c_{13}\right) h r^2 \left(f'\right)^2 \nn\\
\hspace{-0.5cm}
&& +2 f^3 \Big[ a^4 r \left(\left(-c_1+c_2+c_{13}\right) r f' h'+2 h \left(\left(-c_1+c_2+c_{13}\right) r f''+\left(-2 c_1+c_2+2 c_{13}\right) f'\right)\right) \nn\\
\hspace{-0.5cm}
&& +a^2 \left(8 \left(c_2+c_{13}\right) h-4 c_2 r h'\right)-2 \left(-2 c_1+c_2+c_{13}+2 c_{14}\right) h r^2 \left(a'\right)^2 \nn\\
\hspace{-0.5cm}
&& +a^3 \left(-11 c_1+3 c_2+3 c_{13}+8 c_{14}\right) h r^2 a' f'+a \left(c_1-c_2-c_{13}\right) r \left(2 h r a''+a' \left(r h'+4 h\right)\right)\Big] \bigg\}/[4 f^2 r^2 \left(a^2 f-1\right)],
\ea
\end{widetext}
where $c_{ij}=c_i+c_j$.

Introducing the following quantity
\be
\beta := (c_2+c_3-c_4)fa^2+c_{14}\,,
\ee
Then, the coefficients in Eqs.~(\ref{back1})-(\ref{back3}) are 
expressed as
\ba
& &
\alpha_1=-2\beta (a^2 f-1)r^2 f^2\,,\nn \\
& &
\alpha_2=(c_2+c_{13}-2\beta)(a^2 f-1)r^2 a f \,\nn \\
& &
\alpha_3=4 (c_2+c_{13})r^2 f^3 a h\,,\nn \\
& &
\alpha_4=-[(c_2+c_{13}-2\beta)a^2f-2(c_2+c_{13}-\beta)]r^2 a h\,,\nn \\
& &
\alpha_5=2[(2c_2+2c_{13}-3\beta)a^2f+3\beta] r^2 fh\,,\nn \\
& &
\alpha_6=4 (c_2 a^2 f +c_{13}-2\beta) (a^2 f-1) r a f h\,,\nn \\
& & 
\alpha_7=-4 c_2 (a^2 f-1)^2 r af^2\,,\nn \\
& &
\alpha_8=-8\beta (a^2 f-1) rf^2 h\,,\nn \\
& &
\alpha_9=8  (c_2+c_{13}) (a^2f-1)^2af^2 h \,,
\ea
\ba
& &
\beta_1=4\beta (a^2 f-1) r^2 a f^3 \,,\nn \\
& &
\beta_2=-2(c_2+c_{13}-2\beta)(a^2f-1)r^2 a^2 f^2 \,,\nn \\
& &
\beta_3=-4[2(c_2+c_{13})a^2f-\beta] r^2 f^3 h\,,\nn \\
& &
\beta_4=[2(c_2+c_{13}-2\beta)a^4 f^2-8(c_2+c_{13}-\beta)a^2 f
\nn \\
& &\qquad\,
+c_2+c_{13}]r^2 h\,, \nn \\
& &
\beta_5=-4[(2c_2+2c_{13}-3\beta)a^2 f+c_2+c_{13}+\beta]
r^2 a f^2 h\,, \nn \\
& &
\beta_6=-4(a^2f-1)[2+2c_2-2(2c_2-c_{13}+2\beta)a^2 f \nn \\
& &\qquad
+2c_2 a^4 f^2 ]rfh\,, \nn \\
& &
\beta_7=8c_2(a^2 f-1)^2r f^3 a^2\,, \nn \\
& &
\beta_8=16 (c_2+\beta)(a^2 f-1) ra f^3 h\,,\nn \\
& &
\beta_9=4[ 2-\{ 2+4c_2+2c_{13}-2(4c_2+3c_{13})a^2 f\nn \\
& & \qquad
+4(c_2+c_{13})a^4 f^2 \} h] f^2 (a^2 f-1)\,,
\ea
\ba
& &
\mu_1=-4c_2(a^2f-1)r a f^3 \,,\nn \\
& &
\mu_2=2(1+c_2-2c_2 a^2 f)(a^2 f-1)r f \,,\nn \\
& &
\mu_3=4(2c_2+\beta-2c_2 a^2f)r f^3 h\,,\nn \\
& &
\mu_4=[2+3c_2+c_{13}-2(1+c_2+2c_{13}-2\beta)a^2 f \nn \\
& & \qquad
-4c_2 a^4 f^2] rh\,,\nn \\
& &
\mu_5=4(6c_2-7c_2 a^2f-c_{13}+2\beta)r a f^2 h \,,\nn \\
& &
\mu_6=4(a^2f-1)[1+2c_2+c_{13}-3(2c_2+c_{13})a^2 f]f h\,,\nn \\
& &
\mu_7=4(a^2f-1)[1+2c_2+c_{13}-(2c_2+c_{13})a^2 f]f^2\,,\nn \\
& &
\mu_8=-16 (2c_2+c_{13})(a^2 f-1)a f^3 h\,.
\ea
\section*{Appendix B: Coefficients in perturbation equations}
\renewcommand{\theequation}{B.\arabic{equation}} \setcounter{equation}{0}

The coefficients in Eq.~(\ref{oddLag}), which should be 
evaluated on the background (\ref{metric}), are 
\ba
C_1 &=& \frac{(1-c_{13})h}{2 r^2 f}\,,\nn \\
C_2 &=& -\frac{c_{13}b}{2 r^2 f}\,,\nn\\
C_3 &=& -\frac{c_{13}ah}{2 r^2}\,,\nn\\
C_4 &=& \frac{[(2c_{14}-c_{13})(fa'+af')r
+2c_{13}a f]h}{2 r^3 f}\,,\nn\\
C_5 &=& \frac{c_1+c_4 a^2 f}{r^2 f}\,,\nn\\
C_6 &=& \frac{2c_4 a b}{r^2}\,,\nn\\
C_7 &=& \frac{[c_4 (a^2 f-1)-c_1]h}{r^2}\,,\nn\\
C_8 &=& -\frac{[c_{13}(a^2 f-1)+1]h}{2r^4}\,,\nn\\
C_9 &=& \frac{c_{13}b}{r^4}\,,\nn\\
C_{10} &=& \frac{1-c_{13}a^2 f}{2r^4 f}\,,\nn\\
C_{11} &=& \frac{c_{13}a}{r^4}\,,\nn\\
C_{12} &=& -\frac{c_1}{r^4}\,,\nn\\
C_{13} &=& \frac{\lambda}{r^2}-\frac{c_{13}[(rh'+2h-2)f+rh f']}
{2r^4 f} \nn\\
& & -\frac{2c_4 (fa'+f'a)ah}{r^3}\,,
\ea
where $b$ is given by Eq.~(\ref{bso}).


\end{document}